\documentclass[12pt,a4paper]{article}
\usepackage{amsmath,amssymb,epsfig,bm,mathrsfs,feynmp,color}
\newcommand{\be}{\begin{equation}}
\newcommand{\ee}{\end{equation}}
\newcommand{\bea}{\begin{eqnarray}}
\newcommand{\eea}{\end{eqnarray}}

\newcommand{\aap}{A\&A}

\newcommand{\apjl}{{ApJ Lett.}}
\newcommand{\apj}{{ApJ}}
\newcommand{\mnras}{{Mon. Not. RAS}}
\newcommand{\prc}{{Phys. Rev. C}}
\newcommand{\prd}{{Phys. Rev. D}}
\newcommand{\nat}{{Nature}}

\begin{document}
\textwidth=135mm
 \textheight=200mm
\begin{center}
{\bfseries 
Upper critical field  and  (non)-superconductivity of magnetars
}
\vskip 5mm
M. Sinha$^{\dag,\ddag}$, A. Sedrakian$^\dag$
\vskip 5mm
{\small {\it$^\dag$ 
Institute for Theoretical Physics, J. W. Goethe-University, \\D-60438
Frankfurt am Main, Germany
}} \\
{\small {\it$^\ddag$ 
Indian Institute of Technology Rajasthan, Old Residency Road, Ratanada, Jodhpur 342011, India
}}
\\
\end{center}
\vskip 5mm
\centerline{\bf Abstract}

We construct equilibrium models of compact stars using a realistic
equation of state and obtain the density range occupied by the proton
superconductor in strong $B$-fields. We do so by combining the density
profiles of our models with microscopic calculations of proton pairing
gaps and the critical unpairing field $H_{c2}$ above which the  proton
type-II superconductivity is destroyed. We find that magnetars with
interior homogeneous field within the range $0.1 \le B_{16}\le 2$,
where $B_{16} = B/10^{16}$ G, are partially superconducting, whereas
those with $B_{16} > 2$ are void of superconductivity.  We briefly
discuss the neutrino emissivity and superfluid dynamics of magnetars
in the light of their (non)-superconductivity.

\vskip 10mm
\section{\label{sec:intro}Introduction}

The discoveries of the soft $\gamma$-ray repeaters and anomalous X-ray
pulsars during the last decades and their identification with a new
subclass of compact objects - magnetars - focus considerable attention
on the properties of dense matter in strong magnetic fields. The
inferred magnetic fields on the surfaces of these objects are of order
of $B_{s16} \sim 0.01-0.1$. The magnitudes of interior fields in magnetars
are not known, but have been frequently conjectured to be larger than
their surface fields.  Self-gravitating magnetic equilibria containing
type-II superconductors were studied in the low-field limit, where the
$B$-field is not much larger the critical field $H_{c1}\simeq 10^{13}$ G  beyond which
the creation of quantum flux tubes in protonic superconductor is
energetically favorable
~\cite{2008MNRAS.383.1551A,2012MNRAS.419..732L,2013MNRAS.431.2986H}.
These fields are well below the characteristic fields of
magnetars. Non-superconducting self-gravitating equilibria were
studied in the strong field
regime~\cite{1995A&A...301..757B,2012MNRAS.427.3406F} and, as we argue
below, are appropriate for describing magnetars with fields large
enough ($B_{16} > 2$) to destroy superconductivity.

Flux-tube arrays exist in type-II superconductors for $B$-fields up to
the second critical field $H_{c2}$ at which their normal cores touch
each other and superconductivity vanishes.  The microscopic parameters
of the proton superconductor as well as the coherence length and the
magnetic penetration depth depend on the local density and temperature
of the protonic fluid. They have been computed previously in the
zero-temperature limit~\cite{1995ApJ...447..305S}. An elementary
estimate of the $H_{c2}$ field follows from the observation that the
coherence between the members of a Cooper pair will be lost when the
Larmor radius of protons becomes of order of coherence length $\xi_p$ of
Cooper pairs. This field is given by
%----------------------------------------------
\be
H_{c2} = \frac{\Phi_0}{2\pi\xi_p^2} , 
\ee
%----------------------------------------------
where $\Phi_0 = \pi\hbar c/e$ is the flux quantum.

In this contribution we construct equilibrium models of compact stars
using a realistic equation of state and compute the density range
occupied by the proton superconductor by combining the density
profiles of our models with microscopic calculations of proton pairing
gaps, which provide us with the density dependence of the local upper
critical field $H_{c2}$ required to destroy the proton
superconductivity. We then go on to discuss the implications of the
unpairing effect for neutrino cooling of magnetars and their
superfluid dynamics.

\section{Modeling magnetars}\label{sec:nsmodel}

Our first step is to construct a model of a magnetar on the basis of a
realistic equation of state (EoS) of dense nuclear matter. We assume
that the magnetar's interior contains conventional nuclear matter
composed of neutrons, protons and electrons in $\beta$-equilibrium.
As the underlying EoS we take the one derived by
\cite{2004PhLB..595...44Z}, where the interaction between the nucleons
is modelled in terms of the Argonne AV18 two-body interaction combined
with a phenomenological three-body interaction.  We also constructed
models based on relativistic density functional theory
\cite{2013PhRvC..87e5806C} and found results similar to the previous
EoS. The maximum masses of sequences built from these EOS are above the
current observational lower limit 2$M_{\odot}$ on the maximum mass of
any compact star. The EoS  of \cite{2004PhLB..595...44Z}  has a maximum mass
$2.67M_\odot$.

In the core of a neutron star protons pair in the $^1S_0$ channel
because of their relative low density. If the matter is nearly isospin
symmetric at high densities there is also the possibility of $^3D_2$
pairing in neutron-proton matter~\cite{1996NuPhA.604..491A}. We adopt
the $S$-wave gap in proton matter from Ref. \cite{2004PhLB..595...44Z}
which, consistent with the underlying EoS, is based on the Argonne
interaction supplemented by a three-body force. With this input we
solved the Oppenheimer-Volkoff equations to obtain configuration of
given gravitational mass. In the following we will examine two neutron
stars models, one with ``canonical'' 1.4 $M_{\odot}$ mass and another
with the maximum mass $M = 2.67\, M_\odot$.

Because of the density dependence of the microscopic parameters,
notably the gap function, the critical fields $H_{c1}$ and $H_{c2}$
are density dependent, which translates into dependence of these
fields on the radius of the star.  We relate the total nucleonic density to
the proton fraction using fits provided in \cite{2004PhLB..595...44Z}.
\begin{figure}[t]
\begin{center}
\includegraphics[height=6.cm,width=8.5cm]{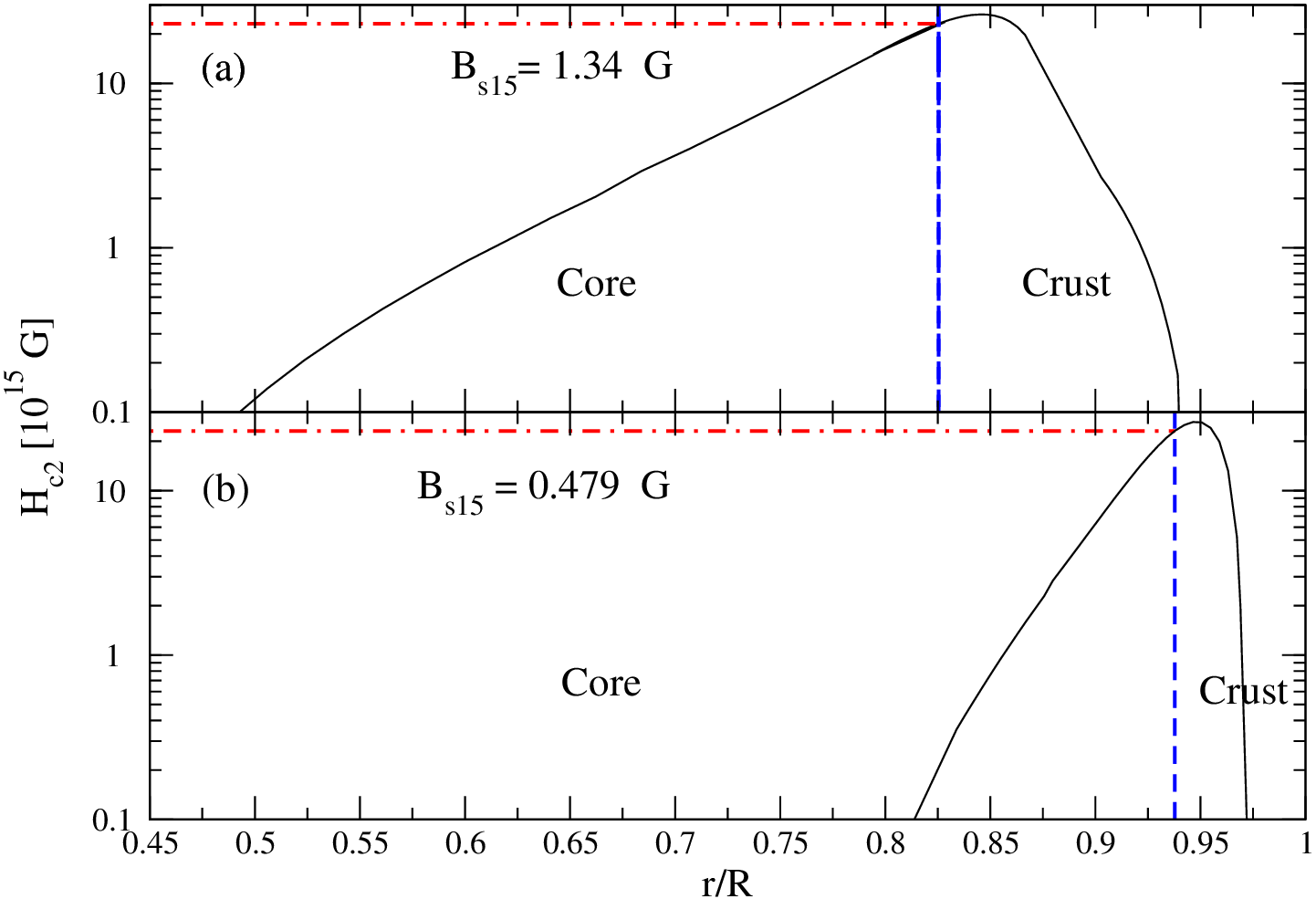}
\caption{Dependence of the critical magnetic field $H_{c2}$ on the
  normalized internal radius $r/R$ of the star for (a) $1.4$
  $M_\odot$, $R = 13.85$ km star and (b) maximum-mass 2.67
  $M_{\odot}$, $R = 11.99$ km star.  The crust core interface
  (vertical dashed line) is located at $R_{\rm core} = 11.43$ km for
  the $1.4$ $M_\odot$ star and at $R_{\rm core} = 11.25$ km for the
  $2.67$ $M_\odot$. The maximal value of $H_{c2}$ in each model is
  attained at the crust-core interface and is indicated by the
  horizontal dash-dotted line; the values of the corresponding surface
  fields are shown in the plot. }
\label{hc2}
\end{center}
\end{figure}
The density dependence of $H_{c2}$ is shown in Fig.~\ref{hc2} for
two models of neutron stars with canonical (1.4 $M_{\odot}$) and
maximal (2.67 $M_{\odot}$) gravitational masses. It is seen that the
maximal value of $H_{c2}$, which is about $2 B_{16}$, is
attained at the crust-core interface and its value drops approximately
linearly with decreasing radius. The field at the crust-core boundary
$H_b$ can be related to the surface field $B_s$ of the star by the
relation $B_s \simeq \alpha_B H_b $ \cite{2013MNRAS.431.2986H}, where
$\alpha_B = \epsilon_b/3$ and $\epsilon_bR$ is the thickness of the
crust, $R$ being the radius of the star.  Assuming the outer boundary
of the core at $0.5n_0$, where $n_0$ is the nuclear saturation
density, we find $\alpha_B = 0.058$ for the 1.4 $M_\odot$ model and $\alpha_B
= 0.021$ for the 2.67 $M_\odot$ model.  
The maximal value of $H_{c2}\simeq 2 B_{16}$ is attained 
at the crust-core boundary and corresponding surface fields are
$B_{s16} = 0.134$ for the 1.4 $M_{\odot}$ model and $B_{s16}
=  0.0479 $ for the maximal mass 2.67 $M_{\odot}$ model.
\begin{figure}[!]
\begin{center}
\includegraphics[height=5.8cm,width=6.6cm]{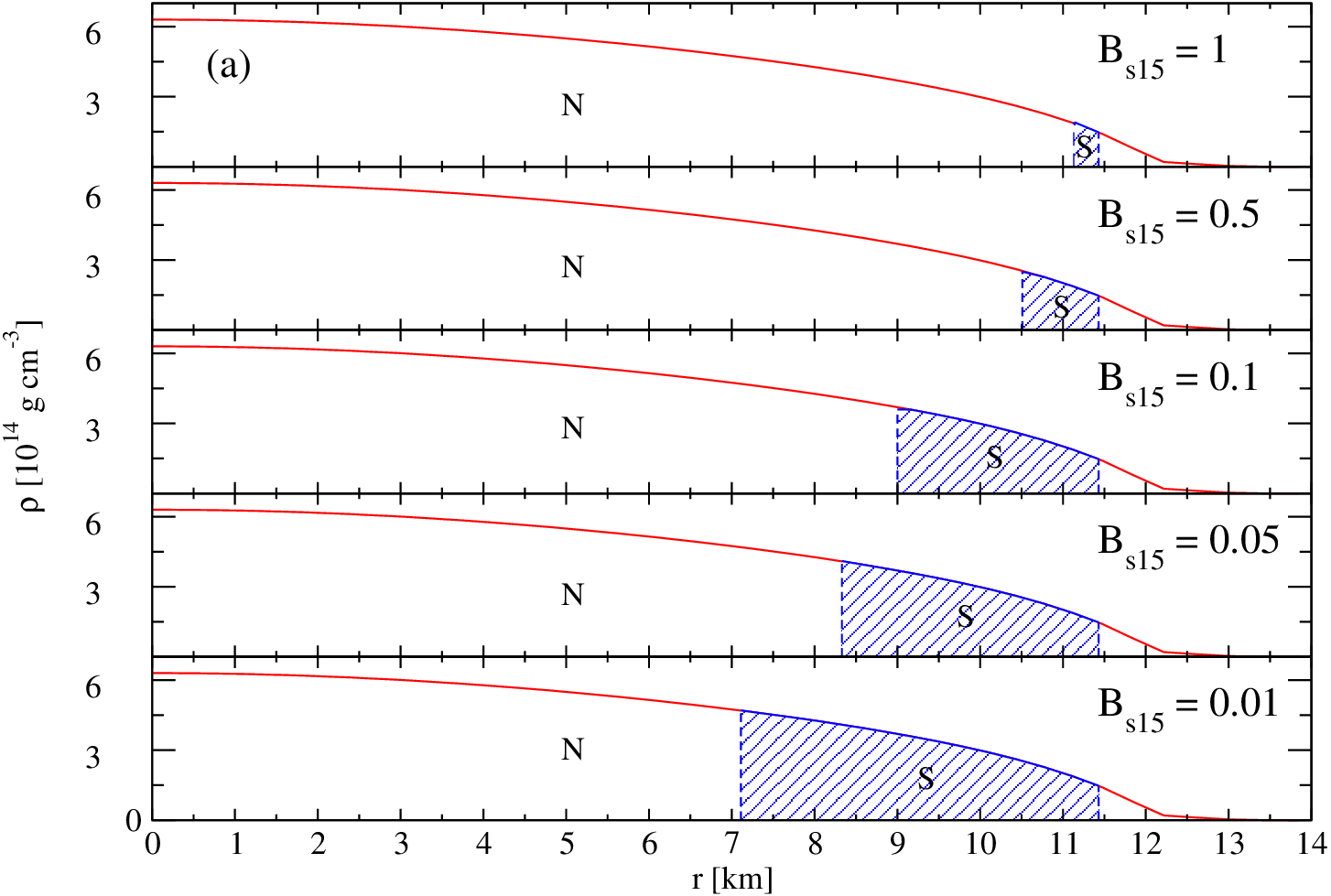}
%\hspace{0.5cm}
\includegraphics[height=5.9cm,width=6.6cm]{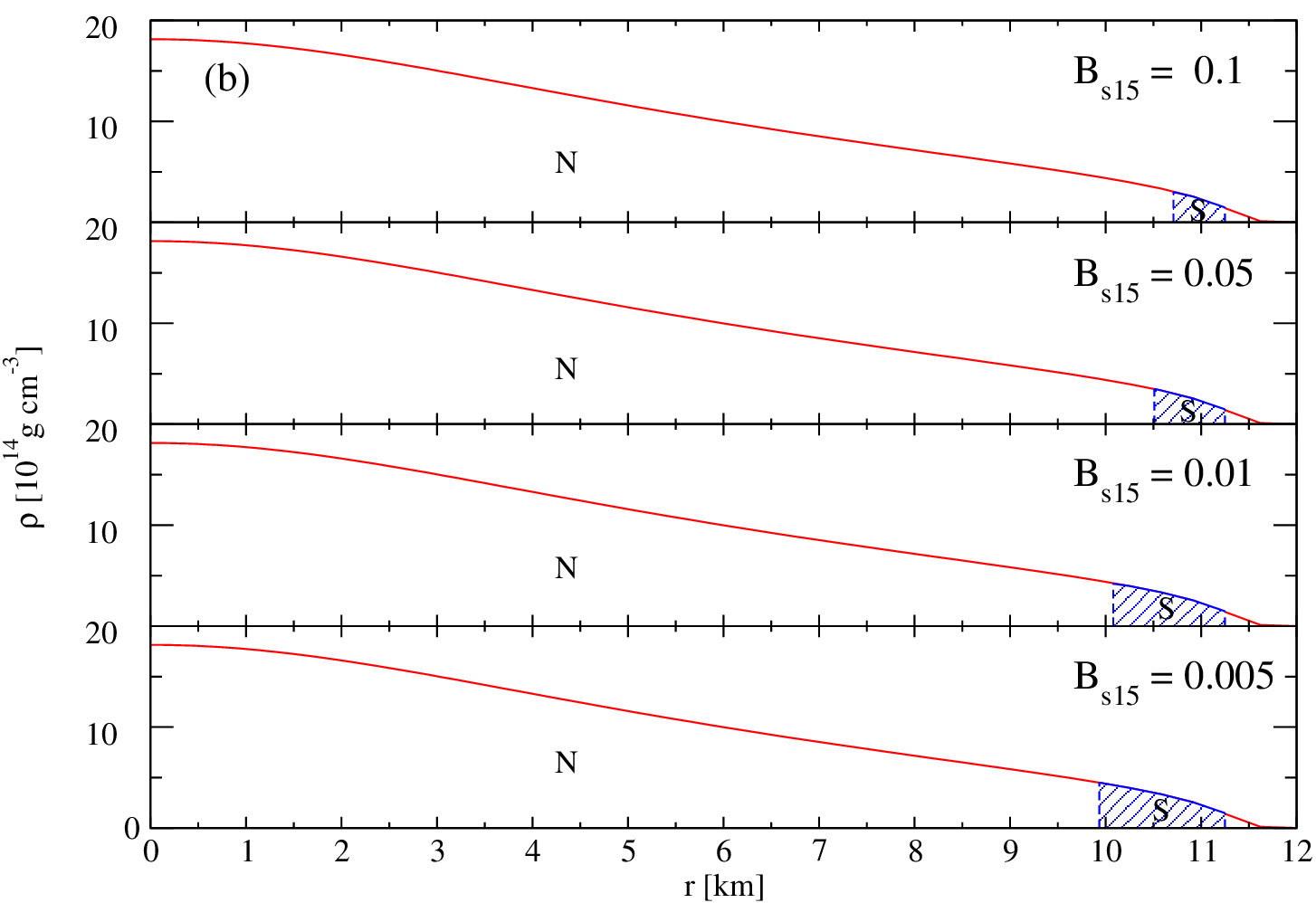}
\caption{Density profiles and the extent of superfluid phases 
(shaded areas) of  $1.4$ $M_\odot$ star (upper panel) and $2.67$
$M_\odot$ star (lower panel) for different values of the surface  field
in units of $10^{15}$ G. The normal and superconducting regions are labeled as
$N$ and $S$.
}
\label{dens1}
\end{center}
\end{figure}
Having established the limiting surface field at which
superconductivity vanishes completely, we now consider decreasing it
down to values characteristic for ordinary neutron stars.
Fig.~\ref{dens1} shows the extend of superconducting phase for
different $B_{s}$, where we related the surface field to the
crust-core boundary field according to the relations above.  It is
seen that with decreasing surface field the superconducting volume
increases and eventually reaches its volume in an ordinary neutron
star. Hence, some magnetars may not have superconducting cores at all,
while some may have, for moderate surface fields, a relatively small
superconducting shell surrounding a core containing unpaired protons.

We conclude this section by stating our main observation: {\it
  intermediate-field magnetars with interior fields $B_{16}\le 2$ are
  partially non-superconducting due to the unpairing effect of the
  magnetic field on the proton superconductivity. Magnetars with
  interior fields $B_{16}\ge 2$ are void of superconductivity.}

\subsection{Neutrino emissivity and heat capacity}
\begin{table}
\begin{tabular}{llcc}
\hline
{\rm Name} & {\rm Process} & {low-$B$} & {high-$B$}\\
\hline
{\rm direct Urca}~~~  & $ n\to p+e+\bar \nu_e$ &  $\times$ & $\checkmark$ \\
{\rm direct bremsstrahlung}~~~  & $ N\to N+\nu_f+\bar \nu_f$ &  $\times$ & $\checkmark$ \\
{\rm PB neutrons}~~~  & $ n+ n\to (nn)+\nu_f+\bar \nu_f$ &  $\checkmark$ & $\checkmark$
\\
{\rm PB protons}~~~  & $ p + p\to (pp)+\nu_f+\bar \nu_f$ &  $\checkmark$ & $\times$
\\
\hline
\hline
\end{tabular}
\caption{Allowed ($\checkmark$) and forbidden ($\times$) process in
  the low and high field neutron stars. Here $N\in n,p$ referes to
  nucleons, $(NN)$ to a Cooper pair.}
\label{table1}
\end{table}
Table~\ref{table1} summarizes the key neutrino emission processes in
the cases of low and high magnetic fields. Strong magnetic fields lift
the kinematical constraint on the direct Urca process, i.e., it can
operate below the Urca threshold on proton
fraction~\cite{1998PhRvD..58l1301B,1999A&A...342..192B}.  As well
known the proton and neutron pairing, which is characterized by the
gaps $\Delta_{n/p}$, cuts the neutrino emission rates, at low
temperatures $T\ll \Delta_{n/p}$ by an exponential factor
$\exp(-\Delta/T)$ for each neutron ($n$) and proton ($p$). If locally
$B> H_{c2}$ the unpairing effect implies that this suppression is
inoperative for protons, i.e., the suppression of the direct Urca
process will be only due to the neutron pairing. Because $\Delta_p \gg
\Delta_n$ the pairing induced suppression differs strongly from the
one expected in the case of superconducting protons.  In $B$-fields an
additional neutrino pair bremsstrahlung channel operates because of
the paramagnetic splitting in the neutron and proton
energies~\cite{2000A&A...360..549V}. The unpairing effect implies that
the direct bremsstrahlung process $p\to p+\nu+\bar\nu$ will not be
suppressed locally if $B> H_{c2}$, which, as in the case of the direct
Urca process, will enhance neutrino emission compared to the low-field
case. A third channel of neutrino emission is the pair-breaking (PB)
neutrino emission from neutron and proton condensates
\cite{1976ApJ...205..541F,2006PhLB..638..114L,2007PhRvC..76e5805S,2008PhRvC..77f5808K}.
The unpairing effect implies that the PB process for protons will be
absent locally if $B> H_{c2}$.

Let us consider, more quantitatively, how the unpairing effect will change
the neutrino luminosity of magnetar's core. Consider first the
  limit $B=0$. The superconducting gap defines a shell in the star
  with a volume $V$.  At non-zero $B$ a part of this shell will become
  normal (see Fig.~\ref{dens1}); we denote the volume of this normal
  shell as $V_N$ and the remainder superconducting volume as $V_S =
  V-V_N$.  The net luminosity of the proton superconducting shell at
  any $B$ can be written as $ L = L_S + L_N$, with $L_{S/N} \equiv
  \bar \epsilon_{S/N} V_{S/N}$, where for the sake of the argument we
  use volume averaged values of the emissivities $ \bar\epsilon_{S/N}
  $. Assuming that superfluidity suppresses the neutrino
  luminosity, its maximum is achieved when the entire shell is normal,
  i.e. $L_{\rm max } = \bar\epsilon_{N} V$. For arbitrary $B$ field
  the luminosity normalized to $L_{\rm max}$ will become
 \be {\cal R} =
  \frac{L}{L_{\rm max}} = 1 -\frac{V_S}{V}
  \left(1-\frac{\bar\epsilon_S} {\bar\epsilon_N}\right) \simeq 1
  -\frac{V_S}{V} , 
\ee 
where the last equality follows in the limit
$\bar\epsilon_S/\bar\epsilon_N\ll 1.$ Because the rightmost
approximate value of $r$ depends only on the volumes of normal and
superconducting shells we can extract its dependence on the
$B_s$-field. For the $M = 1.4 M_{\odot}$ model in Fig.~\ref{dens1} we
find that ${\cal R} (B_{s}/10^{15}\textrm{G})$ has the following
values: ${\cal R} (0.05) = 0.19$, ${\cal R} (0.1) = 0.33$, ${\cal R}
(0.5) = 0.7$, ${\cal R} (1) = 0.9$. Similarly, for the $M = 2.67
M_{\odot}$ model in Fig.~\ref{dens1} we find ${\cal R} (0.01) = 0.1$,
${\cal R} (0.05) = 0.39$, ${\cal R} (0.1) = 0.56$. It is seen that the
increase of the magnetic field leads to an increase of neutrino
luminosity, if unpaired matter emits more neutrinos than the
superconducting one.

Finally, we note that the unpairing effect will enhance the heat
capacity of magnetar's core and, therefore, the magnetar's thermal
relaxation time-scale to a given temperature. In the regions where
locally $B> H_{c2}$, non-superconducting portons will double the heat
capacity, which is mainly provided by relativistic electrons.  The
heat capacity of neutrons forming $S$- and $P$-wave condensates in
negligible.

\subsection{Reheating and rotational response}

The magnetic energy is the key source of internal heating for
magnetars~\cite{1992ApJ...395..250G,2008ApJ...673L.167A}.  Conversion
of magnetic energy into thermal energy depends on the electrical
conducting properties of the fluid core of the star. If protons are
superconducting, the magnetic field is frozen in the flux tubes and
can be changed only if these migrate out of the fluid core. If
however, protons are unpaired by the $B$-field, then the magnetic
field decay produces heat on a time-scale \cite{1969Natur.224..674B}
%-----------------------------------------------------------
\be \tau \propto
\frac{4\pi R^2}{c^2} \sigma 
\ee
%-----------------------------------------------------------
where $\sigma$ is the field-free electrical conductivity, $R$ is a
characteristic scale of the magnetic field. For $R\sim 1$ km, $\tau\sim
10^8$ yr. The decay time-scale for the transverse $B$-field is
substantially reduced in strong fields, where the transverse
conductivity scales as $\sigma_{\perp}^{-1}\sim
B^2$~\cite{1992NuPhA.540..211O}.

The unpairing effect in magnetars has a profound effect on the
superfluid dynamics of its core. In low-field neutron stars the core
dynamics is determined by the interaction of the neutron vortices with
the protonic flux tubes and the electromagnetic interactions of
electrons with this conglomerate. In contrast, non-superconducting
protons will couple to the electron fluid on plasma timescales, which
are much shorter than the hydrodynamical timescales. Therefore, the
core of a magnetar is a two-fluid system with neutron condensate
forming the superfluid component and the porton plus electron fluids
forming the normal component. The neutron superfluid will couple to
the electron-proton plasma by scattering of protons off the neutron
quasiparticles confined in the cores of vortices via the strong
nuclear force. The relaxation time-scale for this process is of order
in the range from several minutes (at the crust-core interface) to a
few seconds (in the deep core) for magnetar periods of order 10 sec
and temperatures of order of $10^8$~K~\cite{Sedrakian:1998uh}.

To summarize we computed the density profiles of the critical field
$H_{c2}$ in realistic models of compact stars and showed that they are
strongly density depend. This implies that magnetars with
approximately homogeneous constant interior $B$-fields are either
completely or partially non-superconducting provided these fields are
by a factor 10 to 15 larger than the observed surface fields of
magnetars.

\section*{Acknowledgements}
We are grateful to S. Lander, H.-J. Schulze, and especially to
I. Wasserman for useful discussion and correspondence.  The work of
M. S. was supported by the Alexander von Humboldt
foundation. A. S. was partially supported by a collaborative research
grant of the Volkswagen Foundation (Hannover, Germany).

%\bibliographystyle{unsrt}
%\bibliography{sup}

\end{document}